\documentclass{ws-jnmp}

\begin{document}


\markboth{LINEARIZATION}{V.~K.~CHANDRASEKAR, M.~SENTHILVELAN and M.~LAKSHMANAN}

%
%

\title{A systematic method of finding linearizing
transformations for nonlinear ordinary differential equations: I. Scalar case
}

\author{V.~K.~CHANDRASEKAR, M.~SENTHILVELAN and M.~LAKSHMANAN}

\address{Centre for Nonlinear Dynamics, Department of Physics,  Bharathidasan
University, \\ Tiruchirapalli - 620 024, India}



\maketitle


\begin{abstract}
In this set of papers we formulate a stand alone method to derive maximal number of linearizing transformations for nonlinear ordinary differential equations (ODEs) of any order including coupled ones from a knowledge of fewer number of integrals of motion.  The proposed algorithm is simple, straightforward and efficient and helps to unearth several new types of linearizing transformations besides the known ones in the literature.  To make our studies systematic we
divide our analysis into two parts.  In the first part we confine our investigations to the
scalar ODEs and in the second part we focuss our attention on a system of two coupled second order ODEs.  In
the case of scalar ODEs, we consider second and third order nonlinear ODEs in detail and discuss the method of
deriving maximal number of linearizing transformations irrespective of  whether it is local or nonlocal type and illustrate the underlying theory with suitable examples. As a
by-product of this investigation we unearth a new type of linearizing transformation in third order
nonlinear ODEs.  Finally the study is extended to the case of general 
 scalar ODEs. We then move on to the study of two coupled second order
nonlinear ODEs in the next part and show that the algorithm brings out a wide variety of linearization transformations. The extraction of maximal number of linearizing transformations in every case is illustrated with suitable examples.

\end{abstract}

\section{Introduction}
\label{int}
The study of linearization of nonlinear ODEs is one of the classic topics but is
yet to be brought to a concise structure.  A systematic
study on this subject had been initiated by Sophus Lie long ago
\cite{olver:1995,Ibragimov1,Steeb}.  In his seminal work he presented the necessary and sufficient conditions for a second order nonlinear ODE to be linearizable under point transformations.
Nevertheless, there has been a revival of interest during the past two decades on linearizing procedures.  In particular,
progress has been made in the following directions.  Durate et al had studied the linearization of a second order nonlinear ODE
by Sundman transformation in which the new independent variable can
be in a nonlocal form \cite{Duarte1,Muriel:09a,Muriel:10}.  Recently, the present authors have proposed certain generalized linearizing
transformations in which the new independent variable is allowed to have derivative terms also besides being nonlocal
\cite{Chand:07}. Apart from these, attempts
have also been made to linearize certain second order nonlinear ODEs by specific nonlocal transformations \cite{Chand:06a}.
As far as the third order nonlinear ODEs are concerned the study on linearization was started by
Lie himself. He investigated the linearization of third order nonlinear ODEs by contact transformations \cite{Ibragimov1}.  The necessary and sufficient
conditions for a third order nonlinear ODE to be linearizable by point transformations were derived by Bocharov et al \cite{Bocharov}, and were later on investigated in detail by
Ibragimov and Meleshko \cite{Ibragimov3}.  The Sundman transformation for third order nonlinear ODE was analysed by Euler et al \cite{Euler1,Euler2}.
Recently the present authors have demonstrated the existence of generalized nonlocal transformations in the case of third order nonlinear ODEs as well \cite{Chand:06}.

The above developments have been essentially concerned with identifying linearizable forms of nonlinear ODEs under various
kinds of transformations.
Throughout this period only very few investigations have been devoted to develop systematic methods to derive linearizing transformations.
As far as our knowledge goes at present Lie symmetry analysis and some ad-hoc methods are being used to derive linearizing transformations.  In fact
both of them have very limited applicability in the case of nonlocal transformations.  Recently, we have proposed
a straightforward procedure to derive linearizing transformations of any type (point transformation, contact transformation, Sundman transformation and generalized linearizing
transformation)
for nonlinear ODEs of any order including coupled ones \cite{Chand:07,Chand:06,Chand:05,Chand:04} starting from an integral of motion associated with the given system.

The main goal of this present set of papers is to penetrate further into the
above mentioned algorithm and bring out multifacted applications of it.  In particular,
we report here for the first time the existence of a new kind of contact transformation for second order nonlinear ODEs which
preserves the order of the equation and in the case of third order nonlinear ODEs a new form of linearizing transformation. Both these results have emerged when we started investigating the role of the other integrals in constructing linearizing transformations for an ODE. Of course in the case of second order nonlinear ODEs once the integral of motion is given one can look for the solution through quadrature. Our aim here is that without quadrature how to deduce the solution from the given integral just by performing algebraic operations on it which is obviously a difficult problem in the case  of higher order ODEs. More interestingly through this ``rewriting" procedure one can get new types of linearizing transformations besides the known ones. We also stress here that the various linearizing transformations, the remaining integrals and the general solution are derived from a known integral.

The algorithm essentially requires one to rewrite the given first integral as a product of perfect derivatives of two functions and to redefine one of them as a new dependent variable and the other as a new independent variable. Interestingly, here we demonstrate that one can rewrite the integral as two perfect derivatives only in a finite number of nontrivial ways (precisely three in the case of second order nonlinear ODEs and two in the case of third order nonlinear ODEs) which in turn gives point transformations and in an infinite number of ways by relaxing the condition that just one of them be a perfect derivative and the others need not be. The latter turn out to be Sundman transformations and generalized linearizing transformations which are infinite in number for
a given ODE. In the case of coupled ODEs, we start with two integrals and rewrite each one of them as products of two functions and investigate the possibility of getting new types of linearizing transformations (in the follow up paper).  Through this attempt we have come up with certain concrete results in the theory of
linearization of nonlinear ODEs.

We organize the rest of the paper as follows. In Sec. \ref{sode}, we consider second order nonlinear ODEs and discuss the method of identifying linearizing transformations, namely point transformation, contact transformation, Sundman transformation and generalized linearizing transformation one by one in detail, starting from an integral. We also derive the general solution for the original equation from the linearized equation for each one of the cases separately. We explain the underlying ideas with suitable examples. In Sec. \ref{tode} we
consider third order nonlinear ODEs and discuss the method of identifying linearizing transformations and general solution in detail. The case of N-th order scalar equations is taken up in Sec. \ref{node}. Finally, we present our conclusion in Sec. \ref{Con}.  Appendix A contains some details about the nonavailability of point/contact transformations from a second integral for third order nonlinear ODEs.
\section{Second order ODEs}
\label{sode}
Let us consider a second order nonlinear ODE
\begin{eqnarray}
\ddot{x}=\phi(t,x,\dot{x}), \;(^{.}=d/dt),
\label{eq01}
\end{eqnarray}
where $\phi$ is a smooth function. Let it admit an integral of motion $I=f(t,x,\dot{x})$ which can be recast in the form \cite{Chand:04,Chand:05}
\begin{eqnarray}
I=f\left(\frac{1}{G(t,x,\dot{x})}\frac{d}{dt}F(t,x)\right),
\label{met01}
\end{eqnarray}
where $F$ and $G$ are functions of their arguments. Such an integral, if it exists, can be found systematically by any one of the recently developed methods like the modified Prelle-Singer procedure \cite{Chand:05}, symmetry based analysis \cite{olver:1995,Ibragimov1,Steeb,Muriel:09a,Muriel:10,Muriel:01,Muriel:09} or Jacobi's last multiplier method \cite{Nucci:07,Nucci:08}, etc. Further, if the function $G(t,x,\dot{x})$ in (\ref{eq01}) is an exact derivative of another function, that is
$G=dz(t,x)/dt$, then (\ref{met01}) can be further simplified to the
form
\begin{eqnarray}
I=f\left(\frac{1}{\frac{dz}{dt}}\frac{dF}{dt}\right)
=f\left(\frac{dF}{dz}\right).
\label{met02}
\end{eqnarray}
Now identifying  the function $w$ as the new dependent variable and $z$
as the new independent variable, that is
\begin{eqnarray}
w = F(t,x),\quad z = \int_o^t G(t',x,\dot{x}) dt',
\label{met05}
\end{eqnarray}
one can rewrite equation (\ref{met02}) in the form $I=f(dw/dz)$. Rewriting this equation as $dw/dz=I_1$, where $I_1$ is a constant (which can also be
treated as the integral of motion) and by simple differentiation one can obtain the free particle equation, $d^2w/dz^2=0$.
The new variables $w$ and $z$ constitute nothing but the linearizing transformation for the given second order nonlinear ODE since they transform the given second order nonlinear ODE into the free particle equation.

With the appropriate selection of the functions $w$ and $z$ one can deduce
all the known linearizing transformations reported in the literature, including point transformation, Sundman transformation and generalized linearizing transformation. In the following we discuss each one of them separately.

\subsection{Point transformations}
\subsubsection{First pair:}
When the function $G$ in (\ref{met05}) is an exact derivative of $t$, that is
$G(t,x,\dot{x})=\bar{G}_{t}+\dot{x}\bar{G}_{x}$, then the linearizing transformation can be simplified to the form $w = F(t,x)$ and $z = \int_o^t G(t',x,\dot{x}) dt'=\bar{G}(t,x)$. As a result one gets the invertible point transformations reported in the literature. In this case we have  the first integral $I_1=dw/dz=\dot{F}/G=(F_{t}+\dot{x}F_{x})/(\bar{G}_{t}+\dot{x}\bar{G}_{x})$. Integrating this equation we get $w=I_1z+I_2$. In terms of $F$ and $\bar{G}$ the latter reads as $F(x,t)=I_1\bar{G}(x,t)+I_2$, where $I_2$ is an arbitrary constant which can be treated as the second integral of motion. Expressing $x$ in terms of $t$ one gets the general solution for the given nonlinear ODE.

A question which now naturally arises is that since the second order nonlinear ODE possesses two integrals, namely $I_1$ and $I_2$, whether one can derive a
second set of linearizing transformation by utilizing the second integral. This can indeed be deduced from the form of the above linearizing transformation itself, as we shall show below.

\subsubsection{Second pair:}
We have, from $w=I_1z+I_2$ and $I_1=\frac{dw}{dz}$,
\begin{eqnarray}
I_2&=w-I_1z=\displaystyle{w-\frac {dw}{dz}z}.\label{met06}
\end{eqnarray}
Now replacing the variables $w$, $z$ and $dw/dz$ in terms of $F$, $G$ and
$\bar{G}$ (vide equation (\ref{met05}) and discussion above),  we get
\begin{eqnarray}
I_2&=
\displaystyle{F-\bigg(\frac{\bar{G}}{G}\bigg)\bigg(\frac{dF}{dt}\bigg)}
=\displaystyle{\frac{1}{G}\bigg(FG-\dot{F}\bar{G} \bigg)}\label{met08}
\end{eqnarray}
which can be rewritten as
\begin{eqnarray}
I_2=-\bigg[\frac{d}{dt}\bigg(\frac{1}{\bar{G}}\bigg)\bigg]^{-1}
\bigg[\frac{d}{dt}\bigg(\frac{F}{\bar{G}}\bigg)\bigg].\label{met09b}
\end{eqnarray}
(In the above we have used the relation $\dot{\bar{G}}=G$).
Equation (\ref{met09b}) can be brought to the form $I_2=dw_1/dz_1$ by
identifying $w_1 = F/\bar{G}$ and $z_1 = 1/\bar{G}$, which in turn yields $d^2w_1/dz_1^2=0$. Thus one gets a second set of linearizing transformation from the integral $I_2$, without actually evaluating its specific form.

\subsubsection{Third pair:}
From the identities $I_2=(F\dot{\bar{G}}-\dot{F}\bar{G})/\dot{\bar{G}}$ and
$I_1=\dot{F}/G$, we can deduce a relation
\begin{eqnarray}
\hat{I_2}=\frac{I_2}{I_1}=\frac{1}{\dot{F}}\bigg(FG-\dot{F}\bar{G} \bigg).
\label{met12}
\end{eqnarray}
Rewriting (\ref{met12}) as $\hat{I_2}=-(d(1/F)/dt)^{-1}[d(\bar{G}/F)/dt]$, we can choose  $w_2 = \bar{G}/F$ and $z_2 = -1/F$, so that $\hat{I_2}= dw_2/dz_2$. Thus we arrive at the free particle equation through a third set of point transformations as well.

\subsubsection{Nonexistence of other pairs}
Now we show the nonexistence of any other linearizing transformation for a second order nonlinear ODE. For this purpose we consider a more general form of the integral given by
\begin{eqnarray}
I=I_1^{r_1}I_2^{r_2}&=&\frac{\dot{F}^{r_1}}{G^{r_1+r_2}}\bigg(FG-\dot{F}\bar{G} \bigg)^{r_2}\nonumber\\
&=&\bigg(\frac{\dot{F}^{\frac{r_1}{r_2}}F}{G^{\frac{r_1}{r_2}}}-\frac{\dot{F}^{\frac{r_1}{r_2}+1}\bar{G}}{G^{\frac{r_1}{r_2}+1}}\bigg)^{r_2},
\label{more01}
\end{eqnarray}
where $r_1$ and $r_2$ are some real numbers.
We find that equation (\ref{more01}) can be recast in the form (\ref{met01}) only for the following three cases: (i) $r_2=0$, $r_1=$ arbitrary (first pair) (ii) $r_1=0$, $r_2=$ arbitrary (second pair) and (iii) $r_1/r_2=-1$ (third pair). For all other cases the right side of (\ref{more01}) becomes nonlinear in $\dot{x}$ and hence it excludes all other possible choices.

As a result one can basically have three different sets of linearizing point transformations for the given second order nonlinear ODE. This is due to the fact that we can derive the linearizing transformations from the integrals, namely $I_1$ and $I_2$, and rewriting them as products of perfect derivative functions. Now each integral gives a linearizing transformation, through the relation, $I_1=dw/dz$ and $I_2=dw_1/dz_1$ and the third pair comes from $I_2/I_1(=(dw_1/dw)(dz/dz_1)=-F^2/\dot{F}(d(\bar{G}/F)/dt))$ which is also nothing but a product of perfect derivatives in the new variables.

The above analysis also reveals the fact that in the case of invertible point transformations one can have the privilege to rewrite the first integral, $I_1=dw/dz$, also in the form
$\hat{I}_1=dz/dw$. In other words one can also treat the new independent variable as a dependent variable and vice-versa. In this sense one can enumerate six sets of linearizing point transformations for a second order nonlinear ODE, though only three are nontrivial.

\subsubsection{Example:}

Let us consider the modified Emden type equation (MEE),
\begin{eqnarray}
\ddot{x}+kx\dot{x}+\frac{k^2}{9}x^3=0,
\label {kps01}
\end{eqnarray}
where $k$ is an arbitrary parameter. A vast amount of literature is
available on the linearization and integrability properties of this equation,
see for example Ref. \cite{Chand:05,mahomed:1985} and references therein.

Equation (\ref{kps01}) admits the following integrals \cite{Chand:05,mahomed:1985}, namely
\begin{eqnarray}
&I_1= -t+\frac{3x}{kx^2+3\dot{x}},\qquad &
I_2=\frac{k}{6}t^2+\frac{1-\frac{k}{3}tx}{\frac{k}{3}x^2+\dot{x}}.
\label{kps04}
\end{eqnarray}
Rewriting the first integral $I_1$ in the form
\begin{eqnarray}
I_1=\frac{1}{\frac{d}{dt}\bigg(\frac{k}{3}t-\frac{1}{x}\bigg)}\left[\frac {d}{dt}\left(\frac{t}{x}
-\frac{k t^2}{6}\right)\right],
\label {kps02a}
\end{eqnarray}
one can get the first pair of linearizing transformation in the form
\begin{eqnarray}
w=F(x,t)=\frac{t}{x}-\frac{kt^2}{6},\quad z=\bar{G}(x,t)=\frac{k}{3}t-\frac{1}{x}, \label{kps03}
\end{eqnarray}
which is nothing but the ones derived by Mahomed and Leach \cite{mahomed:1985}.

Let us now consider the second integral given in (\ref{kps04}) and
rewrite it in the form
\begin{eqnarray}
I_2=\frac{(3-ktx)^2}{kx^2+3\dot{x}}\left[\frac {d}{dt}
\left(\frac{t} {6}-\frac{t}{2(ktx-3)}\right)\right].
\label {kps04a}
\end{eqnarray}
Identifying (\ref{kps04a}) with (\ref{met09b}) one can get the second pair of linearizing transformation of the form
\begin{eqnarray}
w_1=\frac{t} {6}\bigg(1-\frac{3}{ktx-3}\bigg)=\frac{F}{\bar{G}},
\quad z_1=\frac{x}{3-ktx}=\frac{1}{\bar{G}}. \label{kps05}
\end{eqnarray}
One can easily check that in the new variables $(w_1,z_1)$, equation (\ref{kps01})
becomes a free particle equation. The third pair of linearizing transformation can
also be deduced straightforwardly in the form
\begin{eqnarray}
w_2=\frac{2(3-ktx)}{6t-kt^2x}, \quad z_2= \frac{6x}{6t-kt^2x}, \label{kps07}
\end{eqnarray}
from the integral $I_3=I_2/I_1$. As pointed out earlier one may treat the new
independent variable as the dependent variable and vice-versa and enumerate three
more linearizing transformations.

\subsection{Sundman transformation}
It has been shown that the second order nonlinear ODEs can also be linearized by
nonlocal transformations. The simplest example is the Sundman transformation
\cite{Duarte1},
\begin{eqnarray}
w=F(x,t),\quad dz=G(x,t)dt, \label {eq03}
\end{eqnarray}
where $F$ and $G$ are arbitrary smooth functions such that the Jacobian
$J=(\partial (z,w)/\partial (t,x))\neq0$. In the case of Sundman
transformation the first integral in terms of $F$ and $G$ reads as $I_1=(F_{t}
+\dot{x}F_{x})/G(t,x)$. In other words, we have $F_{t}+\dot{x}F_{x}=I_1G(t,x)$ which
upon integration yields
\begin{eqnarray}
F(x,t)=I_1\int G(t,x)dt+C \Rightarrow w=I_1z+C,
\label{spt01ab}
\end{eqnarray}
where $C$ is a constant.
Since $w$ is an explicit function of $x$ and $t$, one can rewrite the latter and obtain
$x=\hat{F}(t,I_1,z,C)$. Substituting this in the expression $dz=G(t,x)dt$ one gets
$dz=\hat{G}(t,I_1,z,C)dt$. We observe that in the case of linearizable equations one can always separate the variables $z$ and $t$ and integrate the resultant equation which in
turn gives $z$ in terms of $t$. Substituting this back in the expression
$x=\hat{F}(t,I_1,z,C)$, where $z$ is now a function of $t$, one gets the general
solution for the given second order nonlinear ODE.

A similar procedure can be adopted for the second integral also to identify the
second pair of Sundman transformation and to obtain the general solution
for the given equation.

\subsubsection{Example:}
To illustrate the ideas in the case of Sundman transformation let us
consider an example given in Ref. \cite{Duarte1},
\begin{eqnarray}
\ddot{x}-\frac{2}{x}\dot{x}^2+\frac{2x}{t^2}=0. \label {ex11}
\end{eqnarray}
Equation (\ref{ex11}) admits two integrals of the form
\begin{eqnarray}
I_1=\frac{\dot{x}t-x}{t^2x^2},\quad I_2= \frac{\dot{x}t^2+2xt}{x^2}.
\label {ex12}
\end{eqnarray}
\subsubsection{Transformation and solution from first pair:}
Rewriting $I_1$ as the perfect derivative of two functions of the form
$I_1=(d(\int x^2 dt)/dt)^{-1}(d(x/t)/dt)$
one can identify the Sundman transformation,
$w=x/t,\quad z= \int x^2 dt$ which in turn transforms equation (\ref{ex11}) to
the free particle equation $d^2w/dz^2=0$ as noted in Ref. \cite{Duarte1}.

Now we follow the procedure given above and deduce the general solution for
this problem. For the present case equation (\ref{spt01ab}) reads as
\begin{eqnarray}
x=I_1zt+Ct. \label {st16}
\end{eqnarray}
Substituting (\ref{st16}) in $dz=t^2(I_1z+C)^2dt$ and integrating it, one gets
$z+C/I_1=\frac{3}{I_1(I_2-I_1^2t^3)}$. Inserting this back in (\ref{st16}), we
obtain the general solution of (\ref{ex11}) in the form
\begin{eqnarray}
x(t)=\frac{3t}{I_2-I_1^2t^3}.\label {st18}
\end{eqnarray}

\subsubsection{Transformation and solution from second pair:}
Now let us consider the second integral given in (\ref{ex12}).
Rewriting this as a product of two perfect derivatives, namely
$I_2=(d(\int x^2 dt)/dt)^{-1})(d(xt^2)/dt)$, one can identify the
second pair of nonlocal transformation of the form
\begin{eqnarray}
w_1=xt^2, \quad z_1= \int x^2 dt. \label{ex17}
\end{eqnarray}
Following the procedure given above one can again deduce the general solution
which in turn exactly matches with (\ref{st18}).

\subsubsection{Infinite sequence of Sundman transformations:}
Next, we identify another interesting result that unlike the fixed number of
linearizing point transformations discussed earlier, one can generate an infinite number of Sundman transformations for a given second order nonlinear ODE. To demonstrate this we rewrite the any one of the integral ($I_1$ or $I_2$) in the form
\begin{eqnarray}
I=\frac{F^n(F_{t}+\dot{x}F_{x})}{F^nG(t,x)}.
\label{spt01b}
\end{eqnarray}
With this choice one can generate a sequence of Sundman transformations of the form
\begin{eqnarray}
&w=F^{n+1}(x,t),\quad &dz=(n+1)F^n(x,t)G(t,x)dt,\quad n\neq-1\nonumber\\
&w=\log{F(x,t)},\quad &dz=\frac{G(t,x)}{F(x,t)}dt, \quad n=-1 \label {spt01c}
\end{eqnarray}
where $n$ is any integer.

Applying this procedure to the present example (\ref{ex11}) one can have an infinite number of Sundman transformations from each integral of motion.
For instance, for the above example (\ref{ex11}), from $I_1$ we get  $w=(x/t)^{n+1}$
and $z=(n+1)\int x^2(x/t)^{n}dt$  and from $I_2$ we obtain $w_1=(t^2x)^{n+1}$ and $z_1=(n+1)\int x^2(t^2x)^{n} dt$ $(n\neq-1)$. For $n=-1$, from $I_1$ we get  $w=\log(x/t)$ and
$z=\int xt dt$ and from $I_2$ we obtain $w_1=\log(t^2x)$ and $z_1=\int 1/(t^2x)dt$.

\subsection{Generalized linearizing transformation}
Our recent studies show that one can also linearize second order nonlinear ODEs
with more generalized nonlocal transformations \cite{Chand:07}. One such generalization
is of the form
\begin{eqnarray}
w=F(x,t),\quad dz=G(t,x,\dot{x})dt, \label {eq06}
\end{eqnarray}
where we have included derivative terms also to define the new independent variable.
In fact, we have shown that a class of equations can be linearized only through this kind of generalized transformation, see for example Ref. \cite{Chand:07}. By proceeding as in the case of the Sundman transformation, one can also deduce these transformations from the integrals of motion.

To construct the general solution for the original equation one may replace the variables $x$ and $\dot{x}$ which appear in $G$ by $t$ so that the resultant equation can be integrated to provide an  expression for $z$ in terms of $t$. Once $w$ and $z$ are known explicitly in terms of $x$ and $t$, just by inverting the free particle equation solution, $w(x,t)=I_1z+C$,
one can arrive at the general solution for the given second order nonlinear ODE.

\subsubsection{Example:} Let us consider the same example which we considered in
the previous case, that is equation (\ref{ex11}) and consider the integral
$(I_2/I_1)$ of the form
\begin{eqnarray}
I_3=\frac{\dot{x}t^4+2t^3x}{\dot{x}t-x},
\label {gt12}
\end{eqnarray}
which in turn yields the generalized linearizing
transformation,
\begin{eqnarray}
w=\frac{1}{xt^2},\quad z= \int \frac{x-\dot{x}t}{x^2t^6}dt. \label{gt14}
\end{eqnarray}
One may note that the new independent variable is in nonlocal form with derivative
terms.  In the new variables equation (\ref{ex11}) is nothing but the free particle
equation.

\subsubsection{Method of finding the solution:}
To obtain $z$ in terms of $t$ we replace $\dot{x}$ and $x$ in terms of $t$ in the
following way. Rewriting (\ref{gt12}) for $\dot{x}$, we get
\begin{eqnarray}
\dot{x}=\frac{x(I_3+2t^3)}{t(I_3-t^3)}.
\label {gt15}
\end{eqnarray}
Now making use the general solution for the free particle equation, $w=I_3z+C$,
that is $1/(xt^2)=I_3z+C$, we get
\begin{eqnarray}
x=\frac{1}{(I_3z+C)t^2}.
\label {gt16}
\end{eqnarray}
Substituting (\ref{gt15}) and (\ref{gt16}) in the second equation in
(\ref{gt14}), we obtain
\begin{eqnarray}
dz=\frac{I_3z+C}{t(I_3-t^3)}dt. \label {gt17}
\end{eqnarray}
Integrating (\ref{gt17}) one gets $z$ in terms of $t$ explicitly. Substituting the
latter now in (\ref{gt16}) we arrive at the general solution which is given in
(\ref{st18}).

\subsubsection{Infinite sequence of generalized linearizing transformations:}

As in the case of the Sundman transformation one can have an infinite
sequence of generalized linearizing transformations for a second order nonlinear
ODE. This fact comes again from the observation that one can rewrite the integrals
of motion in the form
\begin{eqnarray}
I=\frac{F^n(F_{t}+\dot{x}F_{x})}{F^nG(t,x,\dot{x})}.
\label{spg01b}
\end{eqnarray}
Consequently one can deduce a sequence of transformations from $I$ in the form
\begin{eqnarray}
&w=F^{n+1}(x,t),\quad &dz=(n+1)F^n(x,t)G(t,x,\dot{x})dt, \quad n\neq-1\nonumber\\
&w=\log{F(x,t)},\quad &dz=\frac{G(t,x,\dot{x})}{F(x,t)}dt, \quad n=-1.\label {spg01c}
\end{eqnarray}
For the example given in equation (\ref{ex11}), from the third integral (\ref{gt12}), one can
deduce a family of generalized linearizing transformations as
\begin{eqnarray}
&&w=(\frac{1}{xt^2})^{n+1},\quad z= (n+1)\int \frac{x-\dot{x}t}{x^2t^6}(\frac{1}{xt^2})^{n}dt,\quad (n\ne- 1) \label{gt14a}\\
&&w=-\log(xt^2),\quad z= \int \frac{x-\dot{x}t}{xt^4}dt.\quad (n=- 1) \label{gt14b}
\end{eqnarray}

\subsection{Contact Transformations}
Besides the above three types, certain second order nonlinear ODEs can be linearized
through a contact transformation of the form $w=F(t,x,\dot{x})$ and $z=G(t,x,\dot{x})$.
Interestingly this transformation can also be deduced from our procedure.
For illustrative purpose  let us consider the MEE and its first integral again
(vide equation (\ref{kps04})). The first integral associated with the MEE (\ref{kps01})
can also be rewritten in the form
\begin{eqnarray}
I_1= \frac{d}{dt}(\frac{k}{6}t^2+\frac{\dot{x}}{(\frac{k}{3}x^2+\dot{x})^2}).\label {lin05}
\end{eqnarray}
Identifying $w=(k/6)t^2+\dot{x}/((k/3)x^2+\dot{x})^2$ and $z=t$ as the new dependent
and independent variables, respectively, equation (\ref{lin05}) can be brought to the form $I_1= \frac{dw}{dz}$ which in turn leads to the free particle equation $d^2w/dz^2=0$. Integrating the latter, we get $w=I_2+I_1t$, and using the expression
$dw/dz=dw/dt=(k/3)t-kx/(kx^2+3\dot{x})$, we get $(w-\frac{k}{6}t^2)^2/(\dot{w}-\frac{k}{3}t)=(9\dot{x})/(k^2x^2)$ so that
\begin{eqnarray}
\frac{(I_2+I_1t-\frac{k}{6}t^2)^2}{I_1-\frac{k}{3}t}
=\frac{9\dot{x}}{k^2x^2}.\label {lin08}
\end{eqnarray}
Performing a simple integration we arrive at the same solution reported in the literature \cite{Chand:07,Chand:05}.

\subsection{Other possible transformations}
Needless to say one can look for more generalized version of the linearizing transformations dicussed in the above sub-sections. Two such obvious choices one may think of will be
(i) $w=F(t,x,\dot{x})$ and $z=\int G(t,x,\dot{x})dt$ and (ii) $w=\int F(t,x,\dot{x})dt$ and $z=G(t,x,\dot{x})$.  To give support for this type of linearizing transformations, we note that the MEE (\ref{kps01}) can also be linearized to the free particle equation $(d^2w/dz^2=0)$ and third order linear equation $(d^3w/dz^3=0)$ through the nonlocal transformations (i) $w=xe^{\int xdt}$, $z=t$ and (ii) $w=e^{\int xdt}$, $z=t$, respectively \cite{Chand:06a}.  However, in this paper we restrict our attention only to the case in which the new dependent variable is not a nonlocal one.  We will present a detailed account of these results elsewhere.

The discussions and demonstrations presented above clearly show that the method of identifying linearizing transformations from the integral of motion is a versatile one
and can be used for multifaceted applications. In the following we extend the theory to third order nonlinear ODEs.

\section{Third order ODEs}
\label{tode}
The third order nonlinear ODEs of the form $\dddot{x}=\phi(\ddot{x},\dot{x},x,t)$
can be linearized through (i) point transformation \cite{Steeb,Ibragimov:05,
Dmitrieva}, (ii) contact transformation \cite{Bocharov,Ibragimov:05}, (iii) Sundman transformation \cite{Berkovich:00,Euler1,Euler2} and (iv) their generalizations
\cite{Chand:07,Chand:06}. In the following we identify all these transformations from the integrals, linearize the nonlinear ODEs and find their general solutions.

We recall here that in the case of second order nonlinear ODEs we rewrite the
first integral and obtain a relation $dw/dz=I$ which in turn provides the free particle equation by differentiation. In the case of third order nonlinear ODEs while rewriting the
integral one can have two choices: Either express it as $dw/dz=I$ or $d^2w/dz^2=I$, as we
see below. In the first case one ends up with a second order linear ODE, while in the second case a third order linear ODE results.  Nevertheless, we discuss the consequences in both the cases.

\subsection{From third order to linear second order ODEs}
\label{tode1}
Let us assume that the third order nonlinear ODE admits an integral, $I =F(t,x,\dot{x},\ddot{x})$, where $F$ is a function of $t,x,\dot{x}$ and $\ddot{x}$ only. As we did in the second order case let us split the function $F$ as a product of two perfect derivatives and rewrite it as a first order ODE, that is \cite{Chand:06}
\begin{eqnarray}
I=f\left(\frac{1}{G_2(t,x,\dot{x},\ddot{x})}\frac{d}{dt}G_1(t,x,\dot{x})
\right)=f\left(\frac{1}{\frac{dz}{dt}}\frac{dG_1}{dt}\right)
=f\bigg(\frac{dG_1}{dz}\bigg).
\label{the02b}
\end{eqnarray}
Now identifying the function $G_1(t,x,\dot{x})=w$ as the new dependent variable and $z=\int G_2(t,x,\dot{x},\ddot{x}) dt$ as the new independent variable, equation (\ref{the02b}) can be recast in the form $I=f(dw/dz)$. In other words, we have $I_1=dw/dz$, where $I_1$ is a constant, from which we get $d^2w/dz^2=0$. Expressing $w$ and $z$ in terms of the old variables, namely
\begin{eqnarray}
w = G_1(t,x,\dot{x}),\quad z = \int_o^t G_2(t',x,\dot{x},\ddot{x}) dt',
\label{the03}
\end{eqnarray}
one can identify an appropriate linearizing transformations to transform the third
order nonlinear ODE into the free particle equation.  In the following we rewrite the first integral as a second derivative and study the consequence.
\subsection{From nonlinear to linear third order ODEs}
\label{sec3.2}
To transform third order nonlinear ODE into a third order linear ODE we rewrite the
first integral as a perfect second order derivative, that is $I_1=d^2\hat{w}/dz^2$ so that $d^3\hat{w}/dz^3=0$. Since we also have $I_1=dw/dz$, one can get $w=d\hat{w}/dz\Rightarrow \hat{w}=\int w (dz/dt) dt$, so that
\begin{eqnarray}
\hat{w} = \int_o^t G_1(t',x,\dot{x})G_2(t',x,\dot{x},\ddot{x}) dt'
=\int \hat{G}_3(t,x,\dot{x},\ddot{x})dt=G_3(t,x,\dot{x}),
\label{the03a}
\end{eqnarray}
where $G_1$ and $G_2$ are as defined above \cite{Chand:06}. In other words only in the case
$G_1G_2$ is an exact derivative one can rewrite the integral as a perfect second derivative. Then $\hat{w}(t,x,\dot{x})$ and $z$ are the required variables for the third order nonlinear ODE to be linearized to a third order linear ODE.

\subsection{The nature of transformations}
The identified variables $\hat{w}$ and $z$ are rather general and depending upon the
explicit forms of the variables one can get point transformation, contact transformation, Sundman transformation and generalized linearizing transformations. To demonstrate this let us consider the transformation, $\hat{w} = G_3,\; z = \int_o^t G_2 dt'$ (vide equations (\ref{the03a}) and (\ref{the03})), which transforms the given third order nonlinear ODE into a linear equation. Depending upon the forms of the dependent and independent variables one can have any one of the following transformations, that is

(1) point transformation: $\hat{w}=G_3(x,t)$ and $z=\hat{G}_2(x,t)=\int_o^t G_2(t',x,\dot{x}) dt'$,

(2) contact transformation: $\hat{w}=G_3(t,x,\dot{x})$ and $z=\hat{G}_2(t,x,\dot{x})=\int_o^t G_2(t',x,\dot{x},\ddot{x}) dt'$,

(3) Sundman transformation: $\hat{w}=G_3(t,x)$ and $z = \int_o^t G_2(t',x) dt'$,

(4) generalized linearizing transformation: $\hat{w}=G_3(t,x,\dot{x})$ and
$z = \int_o^t G_2(t',x,\dot{x},\ddot{x})dt'$ and

(5) new type of nonlocal transformation: $\hat{w}=\int_o^t \hat{G}_3(t',x,\dot{x},\ddot{x})dt'$ and $z = \hat{G}_2(t,x,\dot{x})dt$.

In the following, we group both the point transformation and the contact transformation together and treat Sundman transformation and generalized linearizing
transformation separately (since in the latter two cases the new independent variable is in integral form) and present our discussion.

\subsection{Point and contact transformations}

\subsubsection{Transformation from $I_1$:}
Using the first choice $(w,z)$ (Sec. \ref{tode1}) one can arrive at the free particle equation from the first integral and consequently obtain $w=I_1z+I_2$,
where $I_1$ and $I_2$ are integration constants (integrals of motion).
Rewriting this expression, $w(t,x,\dot{x})=I_1z(t,x,\dot{x})+I_2$, for $\dot{x}$ and integrating the resultant equation
one gets the general solution for the given third order nonlinear ODE. In the other case (Sec. \ref{sec3.2}), one ends up with $d^3\hat{w}/dz^3=0$ and obtain $\hat{w}=(I_1/2)z^2+I_2z+I_3$ with $I_i$'s, $i=1,2,3$,
 being integration constants. If the transformation is of point type, one can straightforwardly replace $\hat{w}$ and $z$ by $t$ and $x$, and express $x$ in terms of $t$ and obtain the general solution for the original equation.
In the case of contact transformations, in which $\hat{w}$ is an explicit function of $\dot{x}$, one can rewrite the expression
$\hat{w}$ for $\dot{x}$ and substituting the latter into the $w$ expression, that is $w(t,x,\dot{x})=I_1z(t,x,\dot{x})+I_2$, and
rewriting it for $x$ one can obtain the general solution, $x=f(t,I_1,I_2,I_3)$.

Now the question arises, as we observed in the case of second order nonlinear ODEs, whether one
can derive any additional linearizing point/contact transformations from the other integrals $I_2$ and $I_3$ also for the given third order nonlinear ODE. Our analysis shows that one can deduce a second set of linearizing transformations only from the third integral $(I_3)$
which we present in the following and demonstrate that one cannot extract the
linearizing point/contact transformation from $I_2$ in the Appendix.

\subsubsection{Transformation from $I_3$:}
Let us now consider the integral $I_3$, and analyse the consequences.
From the expression $\hat{w}=(I_1/2)z^2+I_2z+I_3$, we find
\begin{eqnarray}
I_3=\hat{w}-\frac{I_1}{2}z^2-I_2z. \label{tmet02}
\end{eqnarray}
Rewriting equation (\ref{tmet02}) in terms of $G_1$ and $G_2$ using equations (\ref{the03}) and  (\ref{the03a}),
we obtain
\begin{eqnarray}
I_3&= \int G_1 G_2 dt-\frac{(\int G_2 dt)^2}{2G_2}\dot{G_1}
-\frac{\int G_2 dt}{G_2}\bigg(G_1G_2-\dot{G_1}\int G_2 dt\bigg)
\label{tmet03}\nonumber\\
&= \int G_1 G_2 dt+\frac{(\int G_2 dt)^2}{2G_2}\dot{G_1}
-G_1\int G_2 dt. \label{tmet04}
\end{eqnarray}
As in the previous case one can rewrite the above expression on the right hand side of (\ref{tmet04}) as a product of two perfect
derivatives to obtain a free particle equation or as
a second derivative to deduce a third order linear ODE.

Let us identify the dependent variable by
rewriting (\ref{tmet04}) as
\begin{eqnarray}
I_3= \frac{-1}{\frac{d}{dt}\bigg(\frac{2}{\int G_2 dt}\bigg)} \frac{d}{dt}
\bigg(G_1-\frac{2\int G_1 G_2 dt}{\int G_2 dt}\bigg). \label{tmet05a}
\end{eqnarray}
Then the new dependent and independent
variables can be chosen as
\begin{eqnarray}
w_{1} = G_1-\frac{2\int G_1 G_2 dt}{\int G_2 dt}\approxeq
\tilde{G_{1}},\quad
z_{1} = \frac{-2}{\int G_2 dt}\approxeq\int{
\tilde{G_{2}}dt},\label{tmet06}
\end{eqnarray}
so that (\ref{tmet04}) can be transformed to
$I_3=dw_{1}/dz_{1}\; \Rightarrow \; d^2w_{1}/dz_{1}^2=0$.
Integrating the latter we obtain $w_{1}=I_3z_{1}+I_2$. We have chosen the integration constant as $I_2$ rather than $I_1$ which can be proved from
equations (\ref{tmet02}), (\ref{tmet04}) and (\ref{tmet06}).

Now we investigate the other possibility, that is to rewrite the
integral $I_3$ as a second derivative of some function. As mentioned earlier this can be implemented
only when the product $\tilde{G_1}\tilde{G_2}$ is a perfect derivative (vide equation (\ref{the03a})). Now we can easily check that
\begin{eqnarray}
\hat{w}_{1} = \int_o^t \bigg(G_1-\frac{2\int G_1 G_2 dt}{\int G_2 dt}\bigg)
\frac{2G_2}{(\int G_2 dt)^2} dt'\label{tmet09a}
\end{eqnarray}
is an exact integral, and (\ref{tmet09a}) can be brought to the form
\begin{eqnarray}
\hat{w}_{1} =2\frac{\int G_1 G_2 dt}{(\int G_2 dt)^2}.
\label{tmet09b}
\end{eqnarray}
Consequently we have $d\hat{w_{1}}/dz_{1}=w_{1}$. Now differentiating this
with respect to $z_1$ and using the identity $dw_{1}/dz_{1}=I_3$ we arrive at
$I_3=d^2\hat{w}_{1}/dz_{1}^2$, which in
turn leads us to $d^3\hat{w}_{1}/dz_{1}^3=0$.
Thus the variables $\hat{w_{1}}$ and $z_{1}$ (vide equations (\ref{tmet06}) and
(\ref{tmet09b})) become a second set of linearizing point/contact transformation for the
given third order nonlinear ODE.

We mention here that one may split equation (\ref{tmet04}) as a product of
two perfect derivatives in certain other decompositions as well.
However, one may not be able to rewrite the integral as a second derivative eventhough it can
be written as a first derivative in the new variables..
To illustrate this point let us rewrite equation (\ref{tmet04}) in a different form, that is
\begin{eqnarray}
I_3=\frac{(\int G_2 dt)^3}{2G_2} \frac{d}{dt}
\bigg(\frac{G_1}{\int G_2 dt}-\frac{\int G_1 G_2 dt}{(\int G_2 dt)^2}\bigg),
\label{ltmet05}
\end{eqnarray}
and identify the new dependent and independent variables from this as
\begin{eqnarray}
w_{2} = \frac{G_1}{\int G_2 dt}-\frac{\int G_1 G_2 dt}{(\int G_2 dt)^2},\quad
z_{2} = \frac{-1}{(\int G_2 dt)^2}.\label{ltmet06}
\end{eqnarray}
Consequently equation (\ref{tmet04}) can be brought to the form
$I_3=dw_{2}/dz_{2}$ which upon integration yields
$w_{2}=I_3z_{2}+I_1$. Here we note that the integration constant turns out to be $I_1$.
Thus one gets a second order free particle equation.

Now let us try to rewrite $I_3$ as a second order derivative. In this case, the
following integral
\begin{eqnarray}
\hat{w}_{2} = 2\int_o^t
\bigg(\frac{G_1G_2}{(\int G_2 dt)^4}-\frac{\int G_1 G_2 dt}{(\int G_2 dt)^5}\bigg)
 dt'
\label{ltmet09}
\end{eqnarray}
should be evaluated explicitly. However, the integration on the right hand side cannot be performed explicitly. This implies that one cannot obtain the linearizing point/contact
transformation by rewriting the integral in the form (\ref{ltmet05}).

Our analysis shows that one can deduce only two sets of point/contact transformations
for the given third order nonlinear ODE, see Appendix for further details. For the
sake of illustrative purpose we present an example for each of the cases (point and contact transformations) separately (Examples 1 and 2) in Table I along with the explicit forms of linearizing transformations from $I_1$ and $I_3$ and the general solution.

\subsection{Sundman transformation and generalized linearizing
transformation}
Since the method of identifying the generalized linearizing transformation and Sundman transformation has been pointed out earlier in the case of second order nonlinear ODEs
let us move straight away to the method of finding the general solution when the new independent variable is in nonlocal form.  In the following, we present a method which is applicable both to generalized linearizing transformation and Sundman transformation.

\subsubsection{Transformation from $I_1$:}
From (\ref{the02b}) considering the integral $I_1$ in the form $I_1=\frac{dw}{dz}$, we obtain
\begin{eqnarray}
G_1(t,x,\dot{x})=I_1\int_o^t G_2(t',x,\dot{x},\ddot{x})dt'+C\Rightarrow
w=I_1z+C. \label{gmet01}
\end{eqnarray}
From equation (\ref{gmet01}) and (\ref{the03a}) we get
\begin{eqnarray}
\hat{w}=G_3(t,x,\dot{x})=I_1\int zdz=\frac{I_1}{2}z^2+Cz+I_3. \label{gmet01a}
\end{eqnarray}
Since $w$ is a function of $t$, $x$ and $\dot{x}$ one can invert equation (\ref{gmet01})
for $\dot{x}$ and obtain
\begin{eqnarray}
\dot{x}=\hat{F}(x,t,z,C,I_1). \label{sol01}
\end{eqnarray}
Substituting this into equation (\ref{gmet01a}) and rewriting the latter in terms of $x$,
one can express $x$ in terms of $t$ and $z$, that is $x=\hat{H}(t,z,I_1,C,I_3)$.
From the first integral we can express $\ddot{x}=K(t,x,\dot{x},C,I_1)=\hat{K}(t,z,I_1,C,I_3)$.
Now substituting the expressions $x$, $\dot{x}$ and $\ddot{x}$ in $dz=G_2(t,x,\dot{x},\ddot{x})dt$ we get an ODE in $z$ and $t$. One can separate the variables $z$ and $t$ and integrate the resultant equation which in turn provides the general solution for the given equation.

\subsubsection{Transformation from $I_3$:}

Rewriting equation (\ref{gmet01a}) in the form $I_3=\hat{w}-(I_1/2)z^2-Cz$ and using equation (\ref{gmet01}) in it we get
\begin{eqnarray}
\hat{I}_3=I_1\hat{w}-\frac{1}{2}w^2, \label{gmet02}
\end{eqnarray}
where $\hat{I}_3=I_1I_3$. Rewriting equation (\ref{gmet02}) in terms of the
variables $G_1$ and $G_2$ we obtain
\begin{eqnarray}
\hat{I}_3&= \frac{1}{G_2}(\dot{G_1}\int G_1 G_2 dt-\frac{1}{2}G_1^2G_2).
\label{gmet03}
\end{eqnarray}

First let us see the possibility of transforming the equation (\ref{gmet03}) into a
free particle equation. With this aim we split equation (\ref{gmet03}) in the form
\begin{eqnarray}
\hat{I}_3= \frac{(\int G_1 G_2 dt)^{\frac{3}{2}}}{G_2} \frac{d}{dt}
\bigg(\frac{G_1}{(\int G_1 G_2 dt)^{\frac{1}{2}}}\bigg). \label{gmet05}
\end{eqnarray}
The new dependent and independent variables from (\ref{gmet05}) can be chosen as
\begin{eqnarray}
w_{1} = \frac{G_1}{(\int G_1 G_2 dt)^{\frac{1}{2}}},\quad
z_{1} = \int \frac{G_2}{(\int G_1 G_2 dt)^{\frac{3}{2}}} dt,\label{gmet06}
\end{eqnarray}
so that equation (\ref{gmet05}) can be brought to the form
$\hat{I}_3=dw_{1}/dz_{1}\Rightarrow d^2w_{1}/dz_{1}^2=0$.
Integrating the latter we get $w_{1}=\hat{I}_3z_{1}+I_2$.

From equation (\ref{the03a}) we get
\begin{eqnarray}
\hat{w}_{1} = \int_o^t \bigg(\frac{G_1G_2}{(\int G_1 G_2 dt)^2}\bigg) dt'=\frac{1}{\int G_1G_2dt}.
\label{gmet09a}
\end{eqnarray}
In other words, we have
$d\hat{w_{1}}/dz_{1}=w_{1}$.
Differentiating the latter with respect to $z$ and using the identity
$dw_{1}/dz_{1}=\hat{I}_3$ one gets $\hat{I}_3=d^2\hat{w}_{1}/dz_{1}^2$,
which in turn gives $d^3\hat{w}_{1}/dz_{1}^3=0$.
Thus the variables $\hat{w_{1}}$ and $z_{1}$ become a second set of Sundman transformation/generalized linearizing
transformation for the given third order nonlinear ODE.

In Table I we present one example for each category (Sundman transformation and generalized linearizing transformation) separately (Examples 3 and 4) along with the explicit forms of linearizing transformations from $I_1$ and $I_3$ and the general solution.

\subsection{A new type of nonlocal transformation}
Upon careful investigation, we also find that one can identify a linearizing transformation in which the new dependent variable is in nonlocal form. This type of linearizing transformation comes out in the case that the function $G_2$ in (\ref{the03}) is an exact derivative of $t$, that is $G_2(t,x,\dot{x},\ddot{x})=\bar{G}_{2t}+\dot{x}\bar{G}_{2x}+\ddot{x}\bar{G}_{2\dot{x}}$, and the right hand side of equation (\ref{the03a}) is not an exact derivative of $t$, that is $\int_o^t G_1(t',x,\dot{x})G_2(t',x,\dot{x},\ddot{x}) dt' =\int G_3(t,x,\dot{x},\ddot{x})dt$. Then the linearizing transformation is modified to the form
$w = h_1(t,x,\dot{x}),\;\hat{w}=\int_o^t h_2(t,x,\dot{x},\ddot{x})dt$ and
$z = h_3(t,x,\dot{x})$. This transformation is different from Sundman transformation/generalized linearizing transformation, since in the present case the dependent variable ($\hat{w}$) is in nonlocal form whereas in the case of Sundman transformation/generalized linearizing transformation the independent variable ($z$)
is in nonlocal form. The method of finding the general solution for this linearizing transformation is as follows:

From the free particle equation we have $w=I_1z+I_2$ so that
\begin{eqnarray}
h_1(t,x,\dot{x})=I_1h_3(t,x,\dot{x})+I_2. \label{new01}
\end{eqnarray}
From equation (\ref{new01}) and (\ref{the03a}) we get
\begin{eqnarray}
\hat{w}=\int w dz=\frac{I_1}{2}z^2+I_2z+I_3. \label{new01a}
\end{eqnarray}
Since $w$ and $z$ are functions of $t$, $x$ and $\dot{x}$ one can invert equation
(\ref{new01}) for $\dot{x}$ and obtain $\dot{x}=\hat{F}(t,x,I_1,I_2)$.
From the first integral we can express $\ddot{x}=K(t,x,\dot{x},I_1)=\hat{K}(t,x,I_1,I_2)$.
Substituting the expressions $\dot{x}$ and $\ddot{x}$ in equation (\ref{new01a}) and rewriting the latter in terms of $x$, one can express $x$ in terms of $t$ and $z$,
that is $x=\hat{H}(t,\hat{w},I_1,I_2)$. Now substituting the expressions $x$, $\dot{x}$
and $\ddot{x}$ in $d\hat{w}=h_2(t,x,\dot{x},\ddot{x})dt =\hat{h}_2(t,\hat{w},I_1,I_2)dt$
we can obtain an ODE in $\hat{w}$ and $t$. One can integrate the resultant equation which
in turn provides the general solution for the given equation. We also present an example
for this category (Example 5) in Table I along with the linearizing transformations
and general solution.

Finally, we can extend the above analysis of linearizing transformations for second
order nonlinear ODEs and third order nonlinear ODEs to n$^{th}$ order nonlinear ODEs
as well. The details are as follows.

\section{n$^{th}$  order ODEs}
\label{node}
In the case of third order ODEs we demonstrated that one can linearize the given equation into a second or third order linear ODE. Extending this idea to fourth order ODE one can demonstrate that given ODE can be linearized to either a second or third or fourth order ODE by appropriately choosing the dependent variable. This result can be extended directly to nth order ODE as discussed below.

Let us assume that the n$^{th}$ order nonlinear ODE, $\frac{d^{n}x}{dt^{n}}=\phi(t,x,\frac{dx}{dt},\ldots,\frac{d^{n-1}x}{dt^{n-1}})$, admits an integral $I =F(t,x,\frac{dx}{dt},\ldots,\frac{d^{n-1}x}{dt^{n-1}})$. Now we split the first integral in the following form
\begin{eqnarray}
I=f\left(\frac{1}{G(t,x,\frac{dx}{dt},\ldots,\frac{d^{n-1}x}{dt^{n-1}})}\frac{d}{dt}H_1(t,x,\frac{dx}{dt},\ldots,\frac{d^{n-2}x}{dt^{n-2}})\right)=f\bigg(\frac{dH_1}{dz}\bigg)
=\frac{dw_1}{dz}
\label{nththe02b}
\end{eqnarray}
so that the new dependent and independent variables, namely
\begin{eqnarray}
w_1 = H_1(t,x,\frac{dx}{dt},\ldots,\frac{d^{n-2}x}{dt^{n-2}}),\quad z = \int_o^t G(t',x,\frac{dx}{dt'},\ldots,\frac{d^{n-1}x}{dt'^{n-1}}) dt',
\label{nththe03}
\end{eqnarray}
transforms the n$^{th}$ order nonlinear ODE into the second order linear equation
$\frac{d^{2}w_1}{dz^{2}}=0$. On the other hand defining the dependent variable in
the form
\begin{eqnarray}
w_{2} &=& \int_o^t H_{1}(t',x,\frac{dx}{dt'},\ldots,\frac{d^{n-2}x}{dt'^{n-2}})G(t',x,\frac{dx}{dt'},\ldots,\frac{d^{n-1}x}{dt'^{n-1}}) dt'\nonumber\\
&=&H_{2}(t,x,\frac{dx}{dt},\ldots,\frac{d^{n-2}x}{dt^{n-2}})
\label{nththe03a}
\end{eqnarray}
one can transform the n$^{th}$ order nonlinear ODE into the third order linear equation
$\frac{d^{3}w_2}{dz^{3}}=0$. Continuing further we can generate a sequence of variables
of the form
\begin{eqnarray}
w_{i} &=& \int_o^t H_{i-1}(t',x,\frac{dx}{dt'},\ldots,\frac{d^{n-2}x}{dt'^{n-2}})G(t',x,\frac{dx}{dt'},\ldots,\frac{d^{n-1}x}{dt'^{n-1}}) dt'\nonumber\\
&=&H_{i}(t,x,\frac{dx}{dt},\ldots,\frac{d^{n-2}x}{dt^{n-2}}).
\label{nththe03b}
\end{eqnarray}
Now combining (\ref{nththe03b}) with $z$ (vide Eq. (\ref{nththe03})), one can transform
the n$^{th}$ order nonlinear ODE into the m$^{th}$ order linear equation, that is
$\frac{d^{m}w_m}{dz^{m}}=0$, $m=3,4, \ldots n$.

Depending upon the nature of the underlying transformations they fall into any one of
the following categories, namely
\begin{enumerate}
\item point transformation: $w_n=H_n(x,t)$ and $z=\hat{G}(x,t)$
\item contact transformation: $w_n$=$H_n(t,x,\frac{dx}{dt},\ldots,\frac{d^{n-2}x}{dt^{n-2}})$ 
and $z$=$\hat{G}(t,x,\frac{dx}{dt},\ldots,\frac{d^{n-2}x}{dt^{n-2}})=\int_o^t G(t',x,\frac{dx}{dt'},\ldots,\frac{d^{n-1}x}{dt'^{n-1}})dt'$ 
\item Sundman transformation: $H_n=H_n(t,x)$ and $z = \int_o^t G(t',x) dt'$
\item generalized linearizing transformation: $w_n=H_n(t,x,\frac{dx}{dt},\ldots,\frac{d^{n-2}x}{dt^{n-2}})$ and
$z = \int_o^t G(t',x,\frac{dx}{dt'},\ldots,\frac{d^{n-1}x}{dt'^{n-1}})dt'$.
\end{enumerate}
\hspace{0.7cm}In the above nonlocality is introduced only in the new independent variables. However one may also consider the cases where the new dependent variable is in nonlocal form. This provides the following new additional linearizing transformations, namely
\begin{enumerate}[(5)]
\item [(5)] $w_n=\int_o^t \hat{H}_n(t',x,\frac{dx}{dt'},\ldots,\frac{d^{n-1}x}{dt'^{n-1}})dt'$ and $z = \hat{G}(t,x,\frac{dx}{dt},\ldots,\frac{d^{n-2}x}{dt^{n-2}})$

\item [(6)]$w_n=\int_o^t \int_o^{t_1} \hat{H}_n(t_2,x,\frac{dx}{dt_2},\ldots,\frac{d^{n-1}x}{dt_2^{n-1}})dt_2 dt_1$ and $z = \hat{G}(t,x,\frac{dx}{dt},\ldots,\frac{d^{n-2}x}{dt^{n-2}})$,
\end{enumerate}
and so on.

\begin{sidewaystable}
\caption{Necessary integrals, linearizing transformations and general solutions of certain third order nonlinear ODEs} \small
\begin{tabular}{|p{.7cm}|p{2.0cm}|p{3.9cm}|p{2.7cm}|p{2.9cm}|p{3.5cm}|p{5.0cm}|}
\hline
S.No & Equation & Integrals & Nature of Transformation & Transformation from $I_1$ & Transformation from $I_3$ & Solution \\
\hline
1& $\dddot{x}=-\frac{3\ddot{x}}{2t}$ &$I_1=t\ddot{x}+\frac{\dot{x}}{2}$ &Point&
$w=\dot{x}t^{\frac{1}{2}}$ \cite{Steeb}&
$w=\dot{x}t^{\frac{1}{2}}-xt^{-\frac{1}{2}}
$ &
\\
& \cite{Steeb}&$I_3=2t\dot{x}-4t^2\ddot{x}-2x$& Transformation &$\hat{w}=x$&$\hat{w}=\frac{-x}{4t}$&$x(t)=2I_1t+2I_2 t^{\frac{1}{2}}+I_3$\\
&&&&$z=2t^{\frac{1}{2}}$&$z=\frac{1}{2}t^{-\frac{1}{2}}$&\\
\hline
2& $\dddot{x}=\frac{x\ddot{x}^3}{\dot{x}^3}$\cite{Bocharov}& $I_1 =\frac{\dot{x}^2-x\ddot{x}}{\dot{x}\ddot{x}}$& Contact  &
$w=\frac{x}{\dot{x}}$ \cite{Bocharov}&
$w=\frac{x}{\dot{x}}-2\frac{t\dot{x}-x}{\dot{x}\log{\dot{x}}}$ &
$x(t)=[\pm\sqrt{I_1^2+I_2^2-2I_1(I_3-t)}$
\\
&&$I_3=\frac{\dot{x}^2-x\ddot{x}}{2\dot{x}\ddot{x}}(\log{\dot{x}})^2$& Transformation &$\hat{w}=\frac{t\dot{x}-x}{\dot{x}}$
&$\hat{w}=2\frac{t\dot{x}-x}{\dot{x}(\log{\dot{x}})^2}$&$\ \ \ \ \ \ -I_1]\times \exp[-[I_1+I_2$\\
&&$\ \ \ \ +\frac{t\dot{x}-x}{\dot{x}}-\frac{x}{\dot{x}}\log{\dot{x}}$&&$z=\log{\dot{x}}$&
$ z=-\frac{2}{\log{\dot{x}}}$&$\ \ \ \ \ \ \mp\sqrt{I_1^2+I_2^2-2I_1(I_3-t)}]/I_1]$\\
\hline
3&$\dddot{x}=\frac{\ddot{x}\dot{x}}{x}$ \cite{Euler2} &$I_1=\frac{\ddot{x}}{x}$&   &
$w=\dot{x},$ \cite{Euler2} &
$w=\sqrt{2}\frac{\dot{x}}{x},
$ &
\\
&&$\hat{I}_3=\frac{\ddot{x}x}{2}-\frac{1}{2}\dot{x}^2$&Sundman &$ \hat{w}=\frac{1}{2}x^2$&$\hat{w}=\frac{-2}{x^2}$,&
$x(t)=\sqrt{2I_2} \cosh{\sqrt{2I_1}(t+I_3)}$\\
&&&transformation&$dz=x dt$&$dz=\frac{2\sqrt{2}}{x^2} dt$&\\
\hline
4& $\dddot{x}=\frac{\ddot{x}^2}{\dot{x}}+\frac{\dot{x}\ddot{x}}{x}$&$I_1=\frac{\dot{x}x}{\ddot{x}}$& Generalized  &
$w=x$ \cite{Chand:06} &
$w=\frac{x}{\sqrt{\dot{x}}}$ &
\\
&\cite{Chand:06}&$I_3=\frac{\dot{x}^2x}{\ddot{x}}-\frac{1}{2}x^2$ &linearizing &$\hat{w}=\dot{x}$&$\hat{w}=\frac{1}{\dot{x}}$&
$x(t)=\sqrt{2I_1I_2}\tan\sqrt{\frac{I_2}{I_1}}(t+I_3)$\\
&&&transformation&$ dz=\frac{\ddot{x}}{x}dt$&$dz=\frac{\ddot{x}}{x\dot{x}^{\frac{3}{2}}} dt$&\\
&&&&&& \\
\hline
5& $\dddot{x}=3\ddot{x}-2\dot{x}$&$I_1=(\dot{x}^2+x\ddot{x}-2x\dot{x})e^{-t}$& New  &
$w=x\dot{x}e^{-2t}$  &
$w=e^{t}(x\dot{x}-x^2)^{-1}$ &
\\
&$+\frac{3\dot{x}^2}{x}-\frac{3\dot{x}\ddot{x}}{x}$&$I_3=\frac{x\ddot{x}+\dot{x}^2-3x\dot{x}+x^2}{\dot{x}^2+x\ddot{x}-2x\dot{x}}e^{t}$
&linearizing &$d\hat{w}=x\dot{x}e^{-3t}dt$&$d\hat{w}=\frac{I_1e^{2t}}{(x\dot{x}-x^2)^3}dt$&
$x(t)=\sqrt{I_2e^{2t}-2I_1e^{t}+I_3}$\\
&\cite{Steeb}&&transformation&$z= -e^{-t}$&$z=(x\dot{x}-x^2)^{-1}$&\\
&&&&&& \\
\hline
\end{tabular}
\label{Table1}
\end{sidewaystable}

\section{Conclusion}
\label{Con}
In the present paper, we have presented a systematic method of finding linearizing transformations starting from an integral of motion associated with a given nonlinear ODE.  To make the
analysis transparent, first we considered a single second order nonlinear ODE.  We then discussed each one of the possible transformations, namely point transformation, Sundman transformation, generalized linearizing transformation and contact transformation.  In some cases they turn out to be infinite in number.  We have
also clarified the method of finding the general solution with the main algorithm for the cases of Sundman transformation, generalized linearizing transformation and contact transformation.
We have extended the same analysis for third order nonlinear ODEs and presented the outcome in detail. Here also we have demonstrated
that our method yields certain new kinds of linearizing transformations as well, besides the  above ones. Finally the underlying ideas have been extended to n$^{th}$ order scalar
ODEs also.

In the present paper we have restricted our attention only on the scalar ODEs only.  The question which
naturally arises is what happens if one extends the algorithm to coupled ODEs.  We consider this hard
question in the follow-up Paper II and report certain interesting results.

\section*{Acknowledgments}
The work of VKC and ML is supported by a Department of Science and Technology (DST), Government of India, IRHPA research project. ML is also supported by a DAE Raja Ramanna Fellowship and a DST -- Ramanna program. The work of MS forms part of a research project sponsored by the DST.

\appendix
\setcounter{section}{1}
\subsection{Absence of point/contact transformations from $I_2$ for third order nonlinear ODEs}
In the case of third order nonlinear ODEs we have demonstrated in Sec. \ref{tode} that one can derive the linearizing transformations only from the first and third integrals. In the following, we
show that from the second integral one cannot extract any point/contact transformation.

Integrating the equation $d^2w_{1}/dz_{1}^2=0$ we get $w_{1}=I_3z_{1}+I_2$ which
can rewritten of the form
\begin{eqnarray}
I_2=w_{1}-I_3z_{1}=w_{1}-\frac {dw_{1}}{dz_{1}}z_{1}.\label{tmet012}
\end{eqnarray}
Substituting (\ref{tmet06}) into (\ref{tmet012}) and rewriting the resultant
equation we get
\begin{eqnarray}
I_2= \frac{2}{G_2} \frac{d}{dt}
\bigg(\frac{G_1\int G_2 dt}{2}-\int G_1 G_2 dt\bigg).
\label{tmet012a}
\end{eqnarray}
The new dependent and independent variables then can be identified as
\begin{eqnarray}
\bar{w}_{1} = \frac{G_1\int G_2 dt}{2}-\int G_1 G_2 dt,\quad
\bar{z}_{1} = \frac{\int G_2 dt}{2}.\label{tmet014}
\end{eqnarray}
In terms of $\bar{w}_{1}$ and $\bar{z}_{1}$ one can go back to the original equation, that is
\begin{eqnarray}
I_2=\frac{d\bar{w}_{1}}{d\bar{z}_{1}}.\label{tmet013}
\end{eqnarray}
Thus one gets a second order free particle equation from $I_2$.

Next we look for the other case. To rewrite the integral as a perfect second order derivative, the integral
\begin{eqnarray}
\hat{\bar{w}}_{1} = \int_o^t \bigg(\frac{G_1G_2\int G_2 dt}{4}
-\frac{G_2\int G_1 G_2 dt}{2}\bigg)
dt'\label{tmet016}
\end{eqnarray}
should be integrated explicitly. While doing so we observe that equation (\ref{tmet016}) cannot be
done so which implies that one cannot extract any linearizing point/contact transformation from $I_2$.

Integrating (\ref{tmet013}) we obtain
\begin{eqnarray}
\bar{w}_{1}=I_2\bar{z}_{1}+I_3,\label{tmet013a}
\end{eqnarray}
in which the integration constant turns out to be $I_3$.

Similarly by rewriting equation (\ref{tmet012a}) in the following form
\begin{eqnarray}
I_2= -\frac{(\int G_2 dt)^2}{G_2} \frac{d}{dt}
\bigg(\frac{G_1}{\int G_2 dt}\bigg),
\label{stmet012a}
\end{eqnarray}
we can identify the new dependent and independent variables as
\begin{eqnarray}
\bar{w}_{2} = \frac{G_1}{\int G_2 dt},\quad
\bar{z}_{2} = \frac{1}{\int G_2 dt}.\label{stmet014}
\end{eqnarray}
Then we can rewrite equation (\ref{stmet012a}) in the form
\begin{eqnarray}
I_2=\frac{d\bar{w}_{2}}{d\bar{z}_{2}}.\label{stmet013}
\end{eqnarray}
Thus one gets a free particle equation from $I_2$. However one cannot extract the point/contact transformation to transform the given equation to linear third order ODE. This is because of the fact that the expression
\begin{eqnarray}
\hat{\bar{w}}_{2} = \int_o^t \bigg(\frac{G_1G_2}{(\int G_2 dt)^3}\bigg)
dt'\label{stmet016}
\end{eqnarray}
cannot be evaluated explicitly which implies that one cannot deduce the linearizing point/contact transformation from $I_2$. Thus one cannot derive point/contact transformation from the second integral fro third order nonlinear ODEs.


\end{document}